\documentclass{llncs}
\usepackage{amsmath, amsfonts, graphicx, booktabs, tabularx, multirow, siunitx, xcolor, soul, bm}
\usepackage{hyperref}
\begin{document}
%
\title{Learning Invariant Feature Representation \\ to Improve Generalization across \\ Chest X-ray Datasets}
\author{Sandesh Ghimire$^{1,2}$, Satyananda Kashyap$^{1}$, Joy T. Wu$^{1}$, \\ Alexandros Karargyris$^{1}$, Mehdi Moradi$^{*}$\\ {$^{*}$\tt\small mmoradi@us.ibm.com}
}

\institute{$^{1}$IBM Almaden Research Center, San Jose, CA;\\  $^{2}$Rochester Institute of Technology, Rochester NY}

\maketitle 
\begin{abstract}
Chest radiography is the most common medical image examination for screening and diagnosis in hospitals. Automatic interpretation of chest X-rays at the level of an entry-level radiologist can greatly benefit work prioritization and assist in analyzing a larger population. Subsequently, several datasets and deep learning-based solutions have been proposed to identify diseases based on chest X-ray images. However, these methods are shown to be vulnerable to shift in the source of data: a deep learning model performing well when tested on the same dataset as training data, starts to perform poorly when it is tested on a dataset from a different source. In this work, we address this challenge of generalization to a new source by forcing the network to learn a source-invariant representation. By employing an adversarial training strategy, we show that a network can be forced to learn a source-invariant representation. Through pneumonia-classification experiments on multi-source chest X-ray datasets, we show that this algorithm helps in improving classification accuracy on a new source of X-ray dataset. 

\keywords{Generalization, Adversarial Training, Learning Theory, Domain Adaptation}
\end{abstract}
\section{Introduction}

Automatic interpretation and disease detection in chest X-ray images is a potential use case for artificial intelligence in reducing the costs and improving access to healthcare. It is one of the most commonly requested imaging procedures not only in the context of clinical examination but also for routine screening and even legal procedures such as health surveys for immigration purposes. Therefore, analysis of X-ray images through several computer vision algorithms has been an important topic of research in the past. Recently, with the release of several large open source public datasets, deep learning-based image classification \cite{rajpurkar2017chexnet,chexpert} has found important applications in this area. The recent outbreak of COVID-19 pandemic and the need for widespread screening has further amplified the need for identification of pneumonia and consolidation findings on X-ray radiographs, as opposed to computed tomography (CT).  

Most of the reported deep learning approaches are trained and tested on the same dataset and/or a single source. This is an unrealistic assumption in the case of medical image analysis with widespread screening applications. In radiology, we can always expect different images coming from different scanners, population, or image settings and therefore we can expect test images are different from the ones used in training. In non-quantitative imaging modalities, such as X-ray, this inconsistency of images across datasets is even more drastic. This is a significant hurdle for the adaptation of automated disease classification algorithms in the practice of radiology. Generalization across X-ray sources is therefore necessary to make deep learning algorithms viable in clinical practice. Recently this has been recognized with the radiology editorial board encouraging testing in \textit{external} test set \cite{bluemke2020assessing}. Some works have tried to answer the question of generalization by intensity normalization and adding Gaussian noise layers to neural networks \cite{KlambauerUMH17} while others use simple ensemble strategy as in \cite{mckinney2020international}.

Drawing ideas from causality and invariant risk minimization \cite{arjovsky2019invariant}, we propose that the key to resolve this issue is to learn features that are invariant in several X-ray datasets, and would be valid features even for the new test cases. \textit{The main contribution of our work is that we enforce feature invariance to source of data by using an adversarial penalization strategy}. We show thus with different X-ray datasets that exhibit similar diseases, but come from different sources/institutions. We have access to four public chest X-ray datasets and validate our method by leave-one-dataset-out experiments of training and testing \cite{rajpurkar2017chexnet,Conference:Wang:CVPR2017,Journal:Johnson:Nature2016}. Given the recent interest in pneumonia like conditions, we chose to target pneumonia and consolidation. \textit{We show that the out of source testing error can be reduced with our proposed adversarial penalization method}. We also perform experiments using Grad-CAM \cite{selvaraju2017grad} to create activation maps and qualitatively evaluate and compare the behavior of the baseline and the proposed method in terms of focus on relevant area of the image.

\section{Related Work}
Earlier works on generalization concentrated on statistical learning theory \cite{vapnik2013nature,bousquet2003introduction}, studying the worst-case generalization bound based on the capacity of the classifier. Later on, differing viewpoints emerged like PAC Bayes \cite{mcallester1999some}, information-theoretic \cite{xu2017information} and stability based methods \cite{bousquet2002stability}. Modern works on generalization, however, find statistical learning theory insufficient \cite{zhang2016understanding} and propose other theories from an analytical perspectives \cite{kawaguchi2018towards}. Our work is quite different from these works. Most of these works are about in-source generalization and assume that data is independent and identically distributed (i.i.d) both in training and testing. We, however, start with the assumption that the training and testing could be from different distributions but share some common, causal features. Based on the principles of Invariant Risk Minimization \cite{arjovsky2019invariant}, we propose the idea that learning invariant features from multiple sources could lead to learning causal features that would help in generalization to new sources.

Another closely related area to our work is that of domain adaptation \cite{sener2016learning,ganin2015unsupervised}, and its application in medical imaging \cite{chen2018semantic}. In a domain adaptation setting, the data is available from source and target domains; but, the labels are available only from the source domain. The objective is to learn to adapt knowledge from the source to predict the label of the target. Although similar in spirit, our work is quite different from domain adaptation in that we do not have target data to adapt to during training. Other ideas of distribution matching like Maximum Mean Discrepancy (MMD) \cite{li2015generative,li2017mmd} are related to our work. In comparison, the adversarial approach is very powerful and easily extendable to more than two sources, which is cumbersome to realize using MMD.

\section{Method}
\textbf{Causation as Invariance:} Following reasoning similar to \cite{arjovsky2019invariant}, we argue that extracting invariant features from many different sources would help the network focus on the causal features. This would help the network generalize to new sources in the future assuming that it would extract causal features from the new X-ray images obtained in the future. 

To force a network to learn invariant features, we propose an architecture as shown in Fig. \ref{fig2} based on adversarial penalization strategy. It has three major components: Feature extractor, Discriminator and Classifier.
Drawing ideas from unsupervised domain adaptation \cite{ganin2015unsupervised}, we train the discriminator to classify which source the image was obtained from just using the latent features extracted by the feature extractor. The discriminator is trained to well identify the source from the features. The feature extractor, however, is trained adversarially to make it very difficult for the discriminator to classify among sources. This way, we force the feature extractor network to extract features from the X-ray images that are invariant across different sources for if there were any element in the latent feature that is indicative of the source, it would be easier for the discriminator to identify the sources. In the end, we expect the feature extractor and discriminator to reach an equilibrium where the feature extractor generates features that are invariant to the sources. Meanwhile, the same features are fed to the disease classifier which is trained to properly identify disease. Hence, the features must be source invariant and at the same time discriminative enough of the disease. Next, we describe three main components of our network.\\
{1. \underline{Feature extractor}:}
The feature extractor is the first component that takes in the input X-ray image and gives a latent representation. In Fig. \ref{fig2}, the feature extractor consists of a Resnet 34 \cite{he2015deep} architecture up to layer 4 followed by a global average pooling layer. \\
{2. \underline{Discriminator}:}
The discriminator consists of fully connected layers that take in features after the global average pooling layer and tries to classify which of the sources the image is obtained from. If adversarial training reaches equilibrium, it would mean that feature representation from different sources are indistinguishable (source invariant).\\
{3. \underline{Classifier}:}
The output of the feature extractor network should not only be source invariant but also be discriminative to simultaneously classify X-ray images according to the presence or absence of disease. In our simple model, we simply use a fully connected layer followed by sigmoid as the classifier.

\subsection{Training}
\begin{figure*}[t]
\centering
\includegraphics[width=1\textwidth]{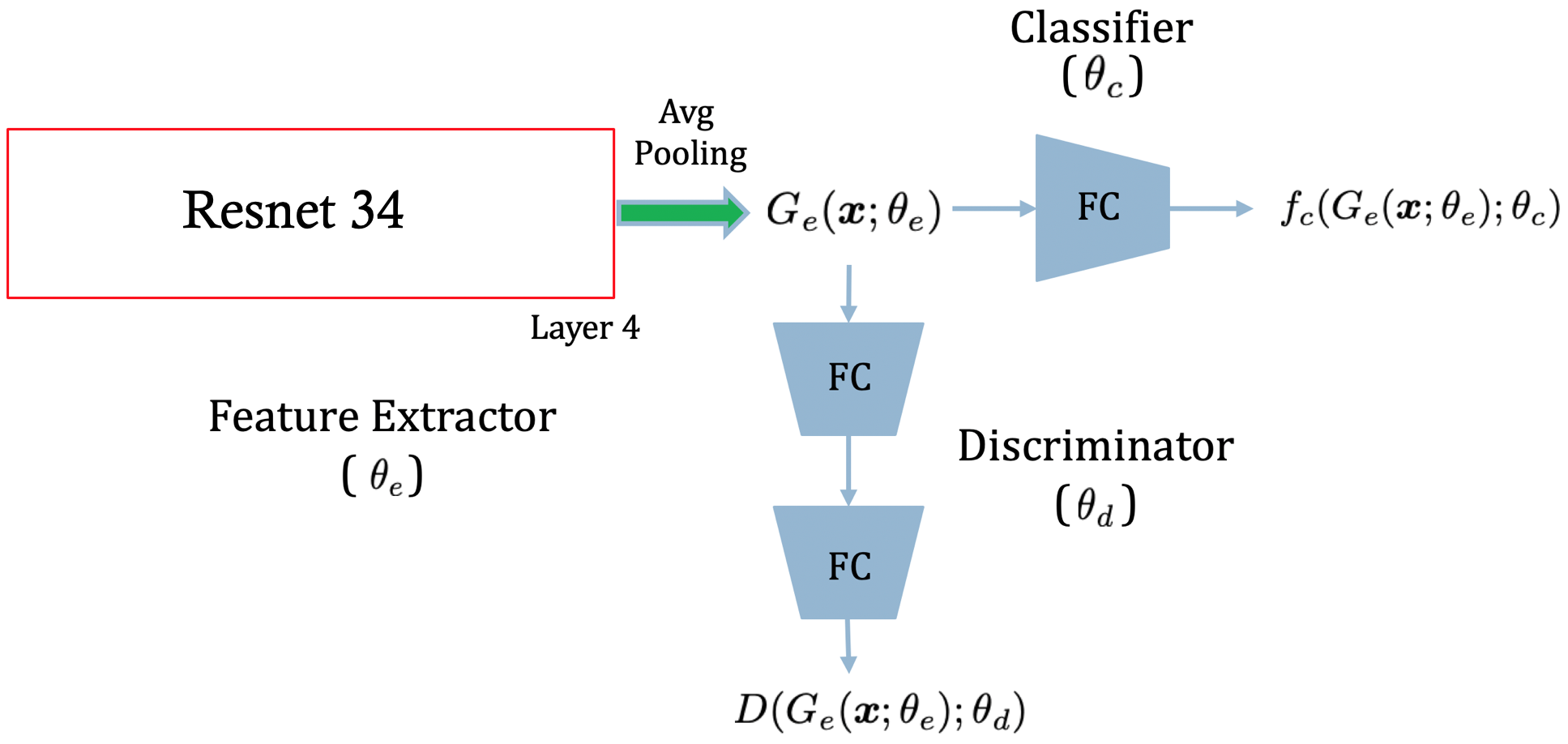}
\caption{\small Proposed architecture to learn source invariant representation while simultaneously classifying disease labels} \label{fig2}
\end{figure*}
From Fig. \ref{fig2}, the disease classification loss and source classification (discrimination) loss are respectively defined as:
\begin{align}
\mathcal{L}_p(\theta_e,\theta_c)=\underset{p(\boldsymbol{x,y})}{E}[\ell_{BCE}(f_c(G_e(x;\theta_e);\theta_c),y)]\\
\mathcal{L}_s(\theta_e,\theta_d)=\underset{p(\boldsymbol{x,y_s})}{E}[\ell_{CE}(D(G_e(x;\theta_e);\theta_d),y_s)]
\end{align}
where $\ell_{BCE}(\hat{y},y)=y\log\hat{y}+(1-y)\log(1-\hat{y})$ is the binary cross entropy loss and similarly $\ell_{CE}(\hat{y},y)=\sum_i (y_s)_i\log(\hat{y}_s)_i+(1-(y_s)_i)\log(1-(\hat{y}_s)_i)$ is the cross entropy loss.
We train extractor, classifier and discriminator by solving following min-max problem.
\begin{align}
\label{eq_cl}
&\hat{\theta_e},\hat{\theta_c}=\underset{\theta_e,\theta_c}{argmin}\hspace{0.2cm} {\mathcal{L}_p(\theta_e,\theta_c)-\lambda \mathcal{L}_s(\theta_e,\hat{\theta}_d)}, \hspace{0.5cm}
\hat{\theta_d}=\underset{\theta_d}{argmin}\hspace{0.2cm} \mathcal{L}_s(\hat{\theta}_e,\theta_d)
\end{align}
It is easy to note that this is a two player min-max game where two players are trying to optimize an objective in opposite directions: note the negative sign and positive sign in front of loss $\mathcal{L}_s$ in eq.(\ref{eq_cl}). Such min-max games in GAN literature are notorious for being difficult to optimize. However, in our case optimization was smooth as there was no issue with stability.

To perform adversarial optimization, two methods are prevalent in the literature. The first method, originally proposed in \cite{goodfellow2014generative}, trains the discriminator while freezing feature extractor and then freezes discriminator to train feature extractor while inverting the sign of loss. The second approach was proposed in \cite{ganin2015unsupervised}, which uses a gradient reversal layer to train both the discriminator and feature extractor in a single pass. Note that the former method allows multiple updates of the discriminator before updating the feature extractor while the latter method does not. Many works in GAN literature reported that this strategy helped in learning better discriminators. In our experiments, we tried both and found no significant difference between the two methods in terms of stability or result. Hence, we used gradient reversal because it was time-efficient.
To optimize the discriminator, it helps if we have a balanced dataset from each source. To account for the imbalanced dataset from each source, we resample data from the source with the small size until the source of the largest size is exhausted. By such resampling, we ensure that there is a balanced stream of data from each source to train the discriminator.

\subsection{Grad-CAM Visualization}
Grad-CAM \cite{selvaraju2017grad} identifies important locations in an image for the downstream tasks like classification. It visualizes the last feature extraction layer of a neural network scaled by the backpropagated gradient and interpolated to the actual image size. 

In this paper, we use Grad-CAM to visualize which location in the X-ray is being attended by the neural network when we train with and without adversarial penalization. We hypothesize that a method that extracts source invariant features should be extracting more relevant features for the disease to be identified, whereas a network which was trained without specific guidance to extract source invariant features would be less focused in the specific diseases and may be attending to irrelevant features in the input X-ray image. Using Grad-CAM, we qualitatively verify this hypothesis.

\section{Datasets and Pneumonia/Consolidation Labeling Scheme}

We used three large scale publicly available datasets for our study - NIH ChestXray14 \cite{Conference:Wang:CVPR2017}, MIMIC-CXR dataset \cite{Journal:Johnson:Nature2016}, and Stanford CheXpert dataset\cite{chexpert}. Further, a smaller internally curated dataset of images originating from Deccan Hospital in India was used.

We are interested in the classification task detecting signs of pneumonia and consolidation in chest X-ray images. Consolidation is a sign of the disease (occurring when alveoli are filled with something other than air, such as blood) whereas Pneumonia is a disease often causing consolidation. Radiologists use consolidation, potentially with other signs and symptoms, to diagnose pneumonia. In a radiology report, both of these may be mentioned. Therefore, we have used both to build a dataset of pneumonia/consolidation. 

We have used all four datasets listed above. The Stanford CheXpert dataset \cite{chexpert} is released with images and labels, but without accompanying reports. 
The NIH dataset is also publicly available with only images and no reports. A subset of 16,000 images from this dataset were de novo reported by crowd-sourced radiologists. For the MIMIC dataset, we have full-fledged de-identified reports provided under a consortium agreement to us for the MIMIC-4 collection recently released \cite{Journal:Johnson:arXiv2019}. For Deccan collection, we have the de-identified reports along with images. For the NIH, MIMIC and Deccan datasets, we used our natural language processing (NLP) labeling pipeline, described below, to find positive and negative examples in the reports; whereas, for the Stanford dataset, we used the labels provided. 

The pipeline utilizes a CXR ontology curated by our clinicians from a large corpus of CXR reports using a concept expansion tool \cite{coden2012spot} applied to a large collection of radiology reports. Abnormal terminologies from reports are lexically and semantically grouped into radiology finding concepts. Each concept is then ontologically categorized under major anatomical structures in the chest (lungs, pleura, mediastinum, bones, major airways, and other soft tissues), or medical devices (including various prosthesis, post-surgical material, support tubes, and lines). Given a CXR report, the text pipeline 1) tokenizes the sentences with NLTK \cite{loper2002nltk}, 2) excludes any sentence from the history and indication sections of the report via key section phrases so only the main body of the text is considered, 3) extracts finding mentions from the remaining sentences, and 4) finally performs negation and hypothetical context detection on the last relevant sentence for each finding label. Finally, clinician driven filtering rules are applied to some finding labels to increase specificity (e.g. "collapse" means "fracture" if mentioned with bones, but should mean "lobar/segmental collapse" if mentioned with lungs). 

Using NLP generated and available labels (for CheXpert), we created a training dataset by including images with a positive indication of pneumonia or consolidation in our positive set and those with no indication of pneumonia or consolidation in the negative set. Table \ref{tab:distribution} lists the number of images from each class for each dataset. These new labels/images will be open-sourced to encourage further research before MICCAI 2020. 

\begin{table*}[t]
\centering
\caption{The distribution of the datasets used in the paper. The breakdown of the Positive (pneumonia/consolidation) and Negative (not pneumonia/consolidation) cases. }

\label{tab:distribution}
\begin{tabular}{@{}lllll@{}}
\toprule
\multicolumn{1}{c|}{\multirow{2}{*}{Leave out Dataset}} & \multicolumn{2}{|c|}{\textbf{Train}} & \multicolumn{2}{c}{\textbf{Test}} \\ \cmidrule(l){2-5} 
\multicolumn{1}{c|}{} & \multicolumn{1}{l|}{Positive} & \multicolumn{1}{l|}{Negative} & \multicolumn{1}{l|}{Positive} & \multicolumn{1}{l}{Negative} \\ \midrule
\multicolumn{1}{l|}{Stanford} & \multicolumn{1}{l|}{15183} & \multicolumn{1}{l|}{123493} & \multicolumn{1}{l|}{1686} & \multicolumn{1}{l}{13720} \\ \midrule
\multicolumn{1}{l|}{MIMIC} & \multicolumn{1}{l|}{83288} & \multicolumn{1}{l|}{49335} & \multicolumn{1}{l|}{23478} & \multicolumn{1}{l}{13704} \\ \midrule
\multicolumn{1}{l|}{NIH} & \multicolumn{1}{l|}{1588} & \multicolumn{1}{l|}{6374} & \multicolumn{1}{l|}{363} & \multicolumn{1}{l}{1868} \\ \midrule
\multicolumn{1}{l|}{Deccan Hospital} & \multicolumn{1}{l|}{50} & \multicolumn{1}{l|}{1306} & \multicolumn{1}{l|}{12} & \multicolumn{1}{l}{379} \\ \midrule
\textbf{Total} & \textbf{100109} & \textbf{180508} & \textbf{25539} & \textbf{29671} \\ \bottomrule
\end{tabular}
\end{table*}
\section{Experiments and Results}
We use four datasets as shown in Table.\ref{tab:distribution}. We use a simple Resnet-34 architecture with classifier as our baseline so that enforcement of invariance through discriminator is the only difference between baseline and proposed method. Experiments using both the architectures use a leave-one-dataset-out strategy: we trained on three of the four datasets and left one out. Each experiment has two test sets: 1)in-source test that draws from only the unseen samples from datasets used for training, 2) out-of-source test set, only including test samples from the fourth dataset that is not used in training. Note that all images from all sources are resized to 512x512.



\begin{table*}[t]
\centering
\caption{The classification results in terms of area under ROC curve from baseline ResNet34 model, and our proposed architecture. Each row lists a leave-one-dataset-out experiment. }
\label{tab:tablecompare}
\begin{tabular}{|l|l|l|l|l|}
\hline
\multirow{2}{*}{\textbf{Leave out Dataset}} & \multicolumn{2}{c|}{\textbf{Baseline}} & \multicolumn{2}{c|}{\textbf{Proposed Architecture}} \\ \cline{2-5}
 & in-source test & out-of-source test  & in-source test & out-of-source test \\ \hline
Stanford                                    & 0.74              & 0.65               & 0.74                   & 0.70                   \\ \hline
MIMIC                                       & 0.80              & 0.64               & 0.80                   & 0.64                   \\ \hline
NIH                                         & 0.82              & 0.73               & 0.71                   & 0.76                   \\ \hline
Deccan Hospital                             & 0.73              & 0.67               & 0.75                   & 0.70                   \\ \hline
\end{tabular}
\end{table*}

\begin{figure*}[ht]
    \centering
    \includegraphics[width=1\textwidth]{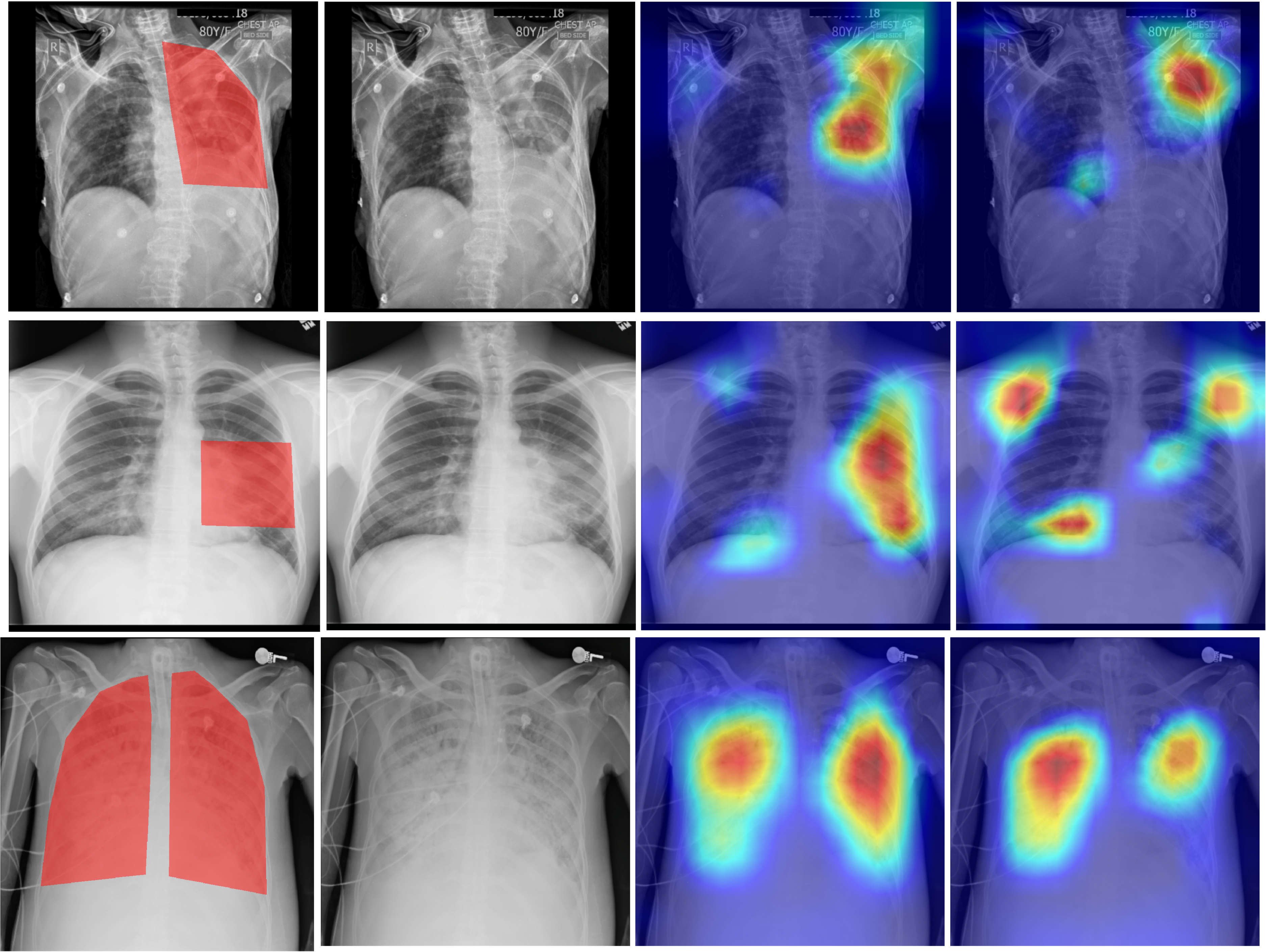}
    \caption{The qualitative comparison of the activation maps of the proposed and the baseline models with the annotation of an expert radiologist. The first column shows the region marked by the expert as the area of the lung affected by pneumonia. The second column shows the original image for reference. The third and fourth columns are the Grad-CAM activation of the proposed and baseline models respectively.}
    \label{fig:gradcam}
\end{figure*}


The results of the classification experiments are listed in Table \ref{tab:tablecompare}. 
We have chosen the area under the ROC curve (AUC-ROC) as the classification metric since this is the standard metric in computer-aided diagnosis. The first observation is that in all experiments, both for baseline and our proposed architecture, the AUC-ROC curve decreases as we move from the in-source test set to the out-of-source test set as expected. However, this drop in accuracy is generally smaller in our proposed architecture. For example, when the Stanford dataset is left out of training, in the baseline method the difference between in-source and out-of-source tests is 0.09 (from 0.74 to 0.65), whereas, in our proposed architecture, the drop in AUC-ROC is only 0.05 (from 0.74 to 0.70). While the performance on the in-source test stays flat, we gain a 5\% improvement in area under the ROC curve, from 0.65 to 0.70 for the out-of-source test.

A similar pattern holds in both the case of NIH and Deccan datasets: in both cases, the drop in performance due to out-of-source testing is smaller for the proposed architecture compared with the baseline classifier. Surprisingly for the NIH dataset, the out-of-source testing results in higher accuracy, which we interpret as heavy regularization during training. In the case of the MIMIC dataset, the performance remains the same for the baseline and the proposed method. 

Fig. \ref{fig:gradcam} shows Grad-CAM visualization to qualitatively differentiate between the regions or features focused by a baseline model and the proposed model while classifying X-ray images. Three positive examples and their activation maps are shown. The interpretation of activation maps in chest X-ray images is generally challenging. However, the evident pattern is that the heatmaps from the proposed method (third column) tend to agree more than the baseline (fourth column) with the clinician's marking in the first column. Furthermore, the proposed method shows fewer spurious activations. This is especially true in row 2 wherein the opacity from the shoulder blades is falsely highlighted as lung pneumonia.

To compare our algorithm with domain generalization approach, we tested on the method in \cite{matsuura2019domain} using pseudo clusters. This methods has the state of the art performance on natural images. On testing with the Stanford leave-out set, the ROC for in-source and out-of-source tests were 0.74 and 0.68 respectively which is slightly below the performance reported here in row 1 of Table 2.

\section{Conclusion and future work}

We tackled the problem of out of source generalization in the context of a chest X-ray image classification by proposing an adversarial penalization strategy to obtain a source-invariant representation. In experiments, we show that the proposed algorithm provides improved generalization compared to the baseline. In the course of this work, we developed labeling methods and applied to the text reports accompanying four datasets to find positive samples for pneumonia/consolidation. These pneumonia/consolidation label lists constitute a new resource for the community and will be released publicly. 

It is important to note that the performance on the in-source test set does not necessarily increase in our method. Mostly it stays flat except in one case, namely the NIH set, where the baseline beats the proposed method in the in-source test. This can be understood as a trade-off between in-source and out-of-source performance induced by the strategy to learn invariant representation. By learning invariant features our objective is to improve on the out-of-source test cases even if in-source performance degrades. 
A possible route for further examination is the impact of the size of the training datasets and left-out set on the behavior of the model.
It is noteworthy that we have kept the feature extractor and classifier components of our current architecture fairly simple to avoid excessive computational cost owing to adversarial training and large data and image size. A more sophisticated architecture might enhance the disease classification performance and is left as future work.

%
%
\bibliographystyle{splncs04}
\bibliography{bibli}

\end{document}